# Kidney and Kidney Tumor Segmentation using a Logical Ensemble of U-nets with Volumetric Validation

Jamie A. O'Reilly[1], Manas Sangworasil[1], Takenobu Matsuura[1]

[1] College of Biomedical Engineering, Rangsit University, Pathumthani, Thailand
`jamie.o@rsu.ac.th`

**Abstract.** Automated medical image segmentation is a priority research area for computational methods. In particular, detection of cancerous tumors represents a current challenge in this area with potential for real-world impact. This paper describes a method developed in response to the 2019 Kidney Tumor Segmentation Challenge (KiTS19). Axial computed tomography (CT) scans from 210 kidney cancer patients were used to develop and evaluate this automatic segmentation method based on a logical ensemble of fully-convolutional network (FCN) architectures, followed by volumetric validation. Data was pre-processed using conventional computer vision techniques, thresholding, histogram equalization, morphological operations, centering, zooming and resizing. Three binary FCN segmentation models were trained to classify *kidney and tumor* (2), and *only tumor* (1), respectively. Model output images were stacked and volumetrically validated to produce the final segmentation for each patient scan. The average F1 score from kidney and tumor pixel classifications was calculated as 0.6758 using preprocessed images and annotations; although restoring to the original image format reduced this score. It remains to be seen how this compares to other solutions.

## 1  Introduction

The objective of this task is to segment kidneys and tumors from axial CT scans automatically, without any manual human intervention. The impetus for this is clear: cases of kidney cancer are notably high, and accordingly, technology for screening medical images may help to ease the workload of radiologists and remove some degree of human error.

Conventional computer vision algorithms can provide reasonably good results on some well-defined medical image processing tasks, with constrained unknown variables. However, these methods have failed to make substantial inroads to clinical practice, presumably due to limited flexibility and requirements for manual fine-tuning. Machine learning approaches based on artificial neural networks currently achieve best performance in a number of image analysis benchmarks, demonstrating promise for future clinical applications. At present, radiologists typically spend a large portion of their working time visually analyzing and annotating medical images, to a large extent manually. This laborious activity can suffer from inter- and intra-observer variability. Thus, automatic methods capable of screening medical images quickly, accu-rately and reliably, would be a valuable clinical tool to improve efficiency. The remainder of this paper describes and evaluates a method submitted to the KiTS19 grand-challenge.

## 2 Method

### 2.1 Data

Axial computed tomography (CT) scans from 210 kidney cancer patients were used, which comprised of 45,424 individual frames; 16,356 containing kidney(s) and 5,696 contained tumor(s) [1]. At the time of writing, a further 90 cases for KiTS19 are yet to be released. Upon inspection, a considerable amount of variation was observed between scans, specifically in terms of frame positioning, size, brightness, perspective, and presence of artifacts (e.g. table). Illustrative samples are shown in the first row of Fig. 1. For these reasons, preprocessing was performed using conventional computer vision techniques to enhance the degree of image uniformity, as described below.

### 2.2 Preprocessing

Images were first resized to 512 x 512 pixels and normalized to pixel values between 0 and 255 (unsigned 8-bit integer format). Histogram equalization and Otsu's thresholding were performed to produce a binary mask. Median and mean filters, with 9 x 9 and 15 x 15 sized kernels, respectively, were convolved with this binary mask to remove noise. A flood-fill algorithm was used to fill holes in the foreground, and then morphological opening was performed with a 99 x 99 kernel to remove smaller foreground objects (specifically aimed at removing the patient table). This binary mask was then multiplied with the image to remove the table artefact and other sources of noise. The resulting image was centered and zoomed to a bounding-box enclosing the body; centering and zooming values were computed then smoothed with a fourth order polynomial function before being applied. Finally, images were down-sized to 256 x 256 pixels. Preprocessed images are shown in the second row of Fig. 1.

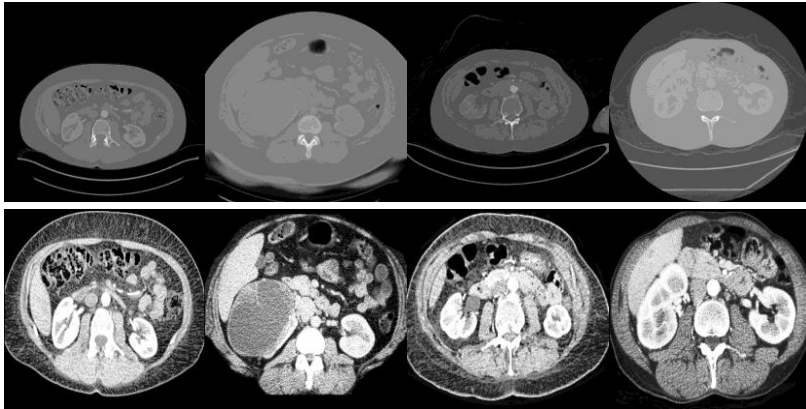

**Fig. 1.** Sample images from the KiTS19 dataset before and after preprocessing. Images in the top row are unedited; although they have been normalized to pixel values 0-255 and stored in .jpeg format. Images in the bottom row have been preprocessed. From left to right: Case 0/Frame 230, Case 56/Frame 30, Case 103/Frame 305, and Case 199/Frame 65.

### 2.3 Image Segmentation

After preprocessing, three two-dimensional fully-convolutional "U-net" models [2, 3] were applied, as illustrated in Fig. 2, to segment images. Two models were trained for predicting kidney-and-tumor regions (one lax and one strict), and one was trained for predicting tumor only. Outputs from the lax kidney-and-tumor classification model were combined with output from the other two models using logical bitwise AND operations. The resulting segmentation masks were added together; representing kidneys with 1, tumors with 2, and everything else equal to 0.

Standard U-net architectures were used, consisting of an encoder stage of nine 2D convolutional layers, four pooling layers, and two 50 % dropout layers; the decoder stage consisted of thirteen convolutional layers and four up-sampling layers. Skip connections across each level of abstraction were applied (four in total), as in the original U-net architecture. The activation functions for input and hidden layers were rectified linear unit, and for the output layer was sigmoid for binary classification.

Three models were trained end-to-end on samples from the data provided. For the strict kidney and tumor model, all CT frames containing only kidneys and tumors were included, along with a random selection of quarter the number of frames without kidneys. This produced 19,479 images in total, with an approximate 4:1 ratio of *containing* to *not-containing* kidneys, used for model training and evaluation. The lax kidney and tumor segmentation model was trained and evaluated using 16,356 images containing the kidneys. For the tumor segmentation model, frames containing tumors were augmented by flipping horizontally to effectively doubling the amount of data for tumor segmentation; a small sample of frames not containing tumors was also included. In each case, an 80:10:10 split was performed to separate data into training, validation, and test sets.

For model training, images were down-sampled to 256 x 256 and fed through the network in randomized mini-batches of twelve at a time. Adaptive momentum optimization (Adam) was employed, with a learning rate of 1 x $10^{-4}$, $\beta_1$ = 0.9, and $\beta_2$ = 0.99. Weighted binary cross-entropy loss was incorporated to account for imbalanced classes, in addition to the selective sampling strategy. Each model was trained for ten iterations, and their outputs were combined as shown in Fig. 2.

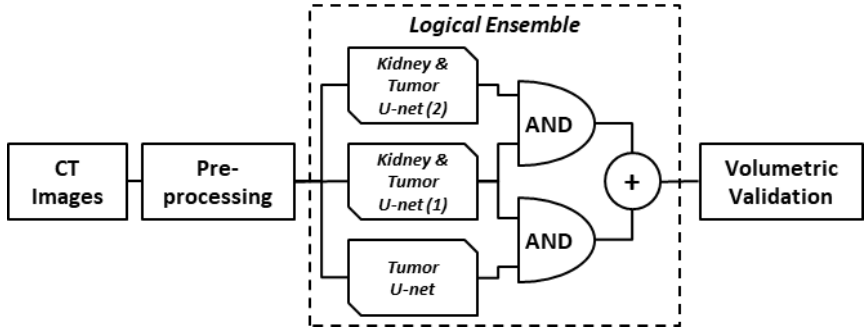

**Fig. 2.** Kidney and tumor segmentation approach block diagram. Kidney & Tumor U-net (1) was lax, while Kidney & Tumor U-net (2) was strict.

### 2.4 Volumetric Validation

Following 2D image segmentation, results for each case were analyzed in three dimensions, with specific criteria for inclusion and exclusion in the final volumetric segmentation. Kidney regions were validated depending upon conditions of volume and position, whereas tumor regions were confirmed broadly based on parameters of geometry and proximity to the confirmed kidney region(s). This approach aims to automatically correct errors in three dimensions that may emerge from stacking segmented images, utilizing basic anatomical knowledge and empirical evidence.

Confirmed kidney regions were limited to volumes above 19,000 normalized units, with center of mass located within the central 60 % of scan depth (i.e. between 20-80 %). Additionally, regions identified as kidney on only one frame were removed. Any regions that failed to meet these criteria were replaced with a label 0, reflecting the background class. Incorrectly assigned tumor region labels were corrected (i.e. replaced with either 0 for background, or 1 for kidney) based on their proximity to the kidneys. The bounding cube for each kidney was determined, and if the tumor bounding cube was found not to intersect with either, then the replacement value was set to 0, for not-kidney; if the bounding cubes were found to overlap, the replacement value was set to 1, for kidney. Tumor regions were confirmed based on volume (> 350 normalized units), position (center of mass located within 80 % of the scan depth), morphology (major axis length > 10, minor axis length > 3; normalized units), consistency (region identified in more than one frame), and sphericity (> 0.29).

### 2.5 Evaluation

In the evaluation, precision, recall, and f1-score are analyzed for each class (background, kidney, and tumor). The formula for these standard performance evaluation metrics are as follows in equations (1-3) below. Please note that these were calculated using pre-processed ground truth labels from the KiTS19 dataset.

$$Precision = \frac{True\ Positive}{True\ Positive + Flase\ Positive} \tag{1}$$

$$Recall = \frac{True\ Positive}{True\ Positive + Flase\ Negative} \tag{2}$$

$$F1\ Score = 2 \times \frac{Precision \times Recall}{Precision + Recall} \tag{3}$$

### 2.6 Software Tools

The following software tools were utilized: Python 3.7.2 (64-bit), Tensorflow 1.13.1, Keras 2.2.4, OpenCV 4.0.0.21, NiBabel 2.3.3, Scikit-Learn 0.20.2, Scipy 1.2.1, and Scikit-Image 0.14.2. Furthermore, model training and evaluation was performed using Google Colaboratory, facilitating access to an NVIDIA Tesla K80 graphical processing unit (GPU) with 12 GB of RAM.

## 3 Results

Results from predicting segmentation volumes using the method portrayed in Figure 2 (with and without volumetric validation) are displayed in Table 1. It may be noted that volumetric validation tended to improve each metric (although it actually decreased recall for the tumor class). It is also apparent that predictions of the tumor class displayed relatively poor precision; all of the other precision and recall values are above 0.8 following volumetric validation.

**Table 1.** Performance evaluations. Metrics of precision, recall, and f1 score are displayed for classes 0-2 from model predictions before and after volumetric validation.

| | Before Volumetric Validation | | | After Volumetric Validation | | |
|---|---|---|---|---|---|---|
| *Metric* | **Class 0** | **Class 1** | **Class 2** | **Class 0** | **Class 1** | **Class 2** |
| **Precision** | 0.9997 | 0.7665 | 0.3449 | 0.9992 | 0.8365 | 0.3964 |
| **Recall** | 0.9960 | 0.8498 | 0.8990 | 0.9975 | 0.8723 | 0.8623 |
| **F1 Score** | 0.9979 | 0.8013 | 0.4510 | 0.9983 | 0.8497 | 0.5019 |

Class 0 = background; Class 1 = kidney; Class 2 = tumor

Sample images from the final segmentation are shown in Figure 4. Please note that these are the same cases and frames presented in Figure 2. It may be noted that the predictions tended to over-estimate tumor regions, either by enlarging correctly la-belled regions or mislabeling the kidneys. This is perhaps preferable to falsely dismiss tumor regions, although neither is ideal. This finding also likely explains the reduced precision for the tumor class, given that this metric penalizes false positives.

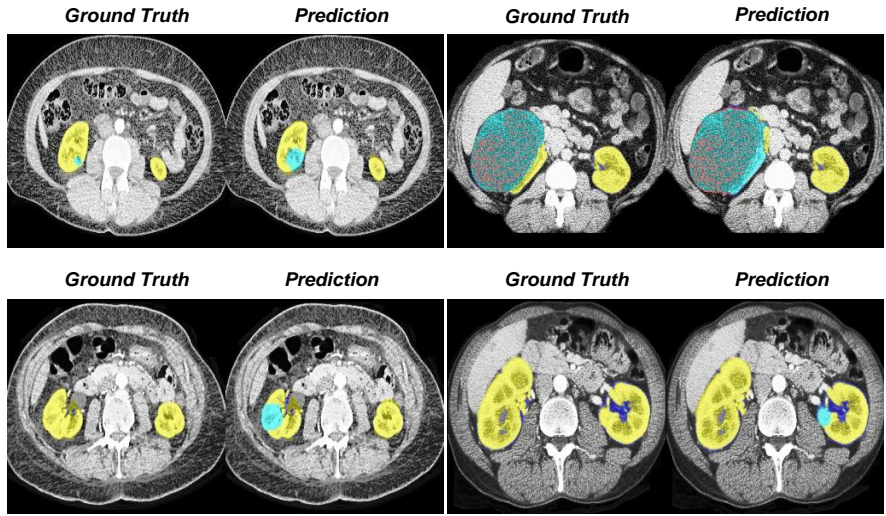

**Fig. 3.** Sample images with corresponding ground truth and predicted labels. Top-left: Case 0/Frame 230. Top-right: Case 56/Frame 30. Bottom-left: Case 103/Frame 305. Bottom-right: Case 199/Frame 65. Kidney regions (class 1) are shaded in yellow; tumor regions (class 2) are shaded in cyan.

To further examine the relationship between kidneys/tumors with the associated segmentation performance, these were plotted in Fig. 4. Some noteworthy cases have been annotated in the relevant plots. Case 15 was found to exhibit unusually large kidneys, perhaps due to renal cysts or tumors. Case 30 presented an extraordinarily large tumor of considerably greater volume than both kidneys combined. Case 88 displayed a large number of extraneous artefacts in images, such as the patient table and blanket, which may have diminished kidney segmentation performance.

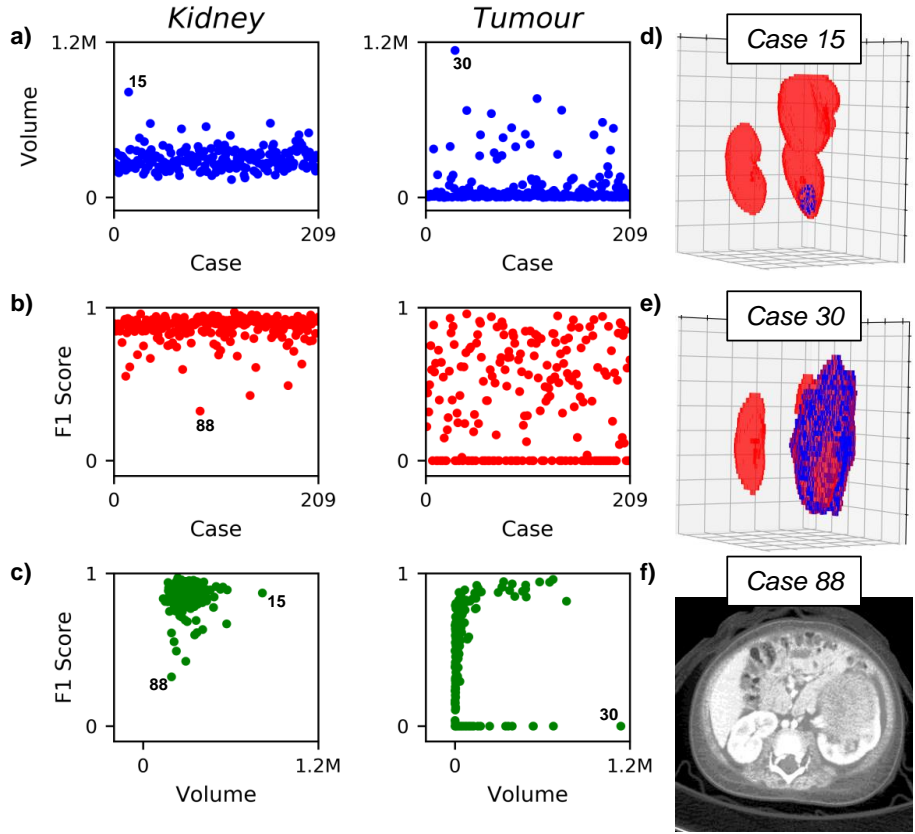

**Fig. 4.** Evaluation of kidney and kidney tumor segmentation performance. Kidney analysis is in the left column and tumor analysis is in the right for (a), (b) and (c). a) Volume per case; volume is plotted in normalized units. b) F1 scores by case. c) F1 scores by volume. Notable cases at the extreme(s) are annotated. d) Volume plot of case 15 showing one abnormally enlarged kidney (red). e) Volume plot of case 30 showing an extremely large tumor (blue). f) Frame from case 88 showing the presence of table and clothing artefacts that were not fully removed by automatic pre-processing.

## 4    Discussion

It can be seen from Table 1 that including volumetric validation improved overall segmentation performance. This is because initial segmentations were performed on 2D images, and no depth information was provided to the machine learning model for it to learn and make predictions about a 3D volume. Potentially, the 3D U-net [4] could have offered an alternative, although this would have involved greater computational demands. As an intermediate solution, it may be possible to include a depth parameter input to the model, so that it may learn the relation between slice depth and kidney/tumor segmentations.

The method described can quite accurately segment the kidneys, with both precision and recall above 0.8. However, for segmenting tumors it displays relatively poor precision (0.4) compared with recall (0.86). The precision metric is more suitable in situations where a false positive is worse (e.g. fraud detection), whereas the recall metric is more appropriate where a false negative is worse (e.g. diagnosing life-threatening illnesses). Hence, the method presented here may be suitable for the application of cancer detection, where the propensity to over-estimate is less deleterious than the tendency to under-estimate the presence of tumors. As a screening procedure followed by verification from a human specialist, this approach may provide efficiency improvements upon current practices. For example, in Figure 4 the non-tumor regions mislabeled as tumors could be identified and corrected by a human operator, perhaps more quickly than without the preceding screening stage. The danger of missing a tumor would still remain, and any efficiency improvements would have to be quantified.

No clear relationship between kidney or tumor volume and segmentation performance can be discerned from Figure 5. As such, it is difficult to identify what cases to focus on to further improve model performance. This may be an instance where more data, or data augmentation, may be of benefit; given that training deep neural networks with more data tends to increase performance. Alternatively, logical ensemble may not be the best design, and rather a multi-class segmentation network may perform better on this task. When the KiTS19 challenge concludes, we will be able to evaluate the different approaches and objectively compare the strengths and weaknesses of each method, which will lead to progress for the field as a whole.

Ultimately, the performance of this method is below the level required for clinical application. This may be improved by further refining the pre-processing protocol to more effectively remove extraneous artefacts, such as the patient table and clothing items, as these clearly impacted the segmentation results for case 88. Equally, measures could be taken in the clinic to avoid unnecessarily introducing artefacts into images destined for automated analysis. Moreover, changes in neural network design and data utilization may also bring improvements.

## Acknowledgments

Credit is due to the outstanding community of bloggers, enthusiasts, and researchers who contribute to online content that enables many others to get involved in machine learning. We also give thanks to Rangsit Research Institute for supporting this work [grant no. 90/2561].

## Code Availability

Code for this project is available from https://github.com/Aj-Jamie/KiTS_Code.